# First-principles investigations of the electronic, magnetic and thermoelectric properties of VTiRhZ (Z= Al, Ga, In) Quaternary Heusler Alloys

## Hind Alqurashi[1,2*], Raad Haleoot[3], and Bothina Hamad [4,5]

[1] Materials Science and Engineering, University of Arkansas, Fayetteville, AR 72701, USA
[2] Physics Department at the College of Science, Al Baha University, Al Baha 65527, Saudi Arabia
[3] Department of Physics at the College of Education, University of Mustansiriyah, Baghdad 10052, Iraq
[4] Department of Physics University of Arkansas, Fayetteville, AR 72701, USA
[5] Physics Department, The University of Jordan, Amman-11942, Jordan

## Abstract

Calculations using density functional theory (DFT) were performed to investigate the structural, dynamical, mechanical, electronic, magnetic, and thermoelectric properties of VTiRhZ (Z = Al, Ga, In) alloys. The most stable structure of these alloys was found to be the type-I configuration. Using GGA-PBE functional, VTiRhGa, and VTiRhIn alloys are predicted as half-metallic ferromagnets with a 100% spin-polarization and a total magnetic moment of $3\mu_B$, which is promising for spintronic applications. The thermoelectric properties and lattice thermal conductivity of VTiRhZ alloys were obtained using the Boltzmann transport theory within the constant relaxation time and Slack equation, respectively. The figure-of-merit ($ZT$) values of VTiRhAl, VTiRhGa, and VTiRhIn alloys were found to be 0.96, 0.88 and 0.64, respectively, which are promising for future thermoelectric applications.

*Keywords*: *Quaternary Heusler alloys, DFT, half-metallic, magnetic properties, mechanical properties, thermoelectric properties*

* Corresponding author: halquras@uark.edu

## 1. Introduction

Recently, Heusler alloys have received a considerable attention as promising candidates in spintronics and thermoelectric applications. There are four categories of Heusler alloys that include full Heusler alloys (FHA), Half-Heusler alloys (HHA), Quaternary-Heusler alloys (QHAs) and Inverse Heusler alloys. The FHAs have the chemical formula ($X_2YZ$) with four interpenetrating cubic lattices, where X and Y are transitional metals (Y could be a rare-earth atom) and Z is an *s-p* element (main group element). Here, X atoms occupy the 8c (1/4, 1/4, 1/4) Wyckoff position with $T_d$ symmetry, whereas the Y and Z atoms occupy 4a (0, 0, 0) and 4b (1/2, 1/2, 1/2) Wyckoff positions, respectively, with the $O_h$ symmetry [1] . When the valence numbers of Y are less than X, the resulting alloy has a space group of $Fm\overline{3}m (no. 225)$ and it crystallizes in the $L_{21}$ structure [1]. Nonetheless, they have the so-called XA crystal structure when the valence number of X element are less than that of Y, which are known as the inverse Heusler alloys with the space group $F\overline{4}3m$ [2], [3]. Here, the X element occupy 4a (0, 0, 0) and 4c (1/4, 1/4, 1/4) Wyckoff positions, whereas the Y and Z elements occupy 4a (0, 0, 0) and 4b (1/2, 1/2, 1/2) Wyckoff positions, respectively. The HHAs, however, have the chemical composition (XYZ) forming three interpenetrating cubic lattices and one vacant lattice. They have a space group of $F\overline{4}3m$ with the $C1_b$ crystal structure [4]. The QHAs, however, have the XX′YZ chemical formula with Y-type or LiMgPdSb-type crystal structure and a space group of $F\overline{4}3m$  (#216) [5].

Heusler alloys could be metals, semiconductors, spin gapless semiconductors, or half-metallic materials. Most of these alloys demonstrate half-metallic electronic properties. These half-metallic materials (HMMs) have a unique electronic structure where one spin channel show a metallic behavior, whereas the other spin channel is semiconducting.  This behavior of HMMs leads to an excellent spin-polarization (100%) near the Fermi energy, which maximizes the efficiency of magneto-electronic devices, such as giant magnetoresistance and tunneling magnetoresistance [6]. In 1983, the half-metallic behavior of Heusler alloys such as NiMnSb was predicted by Groot *et al.* [7]. From then on, a lot of HMMs based on Heusler alloys have been investigated using first-principal calculations, where several of them have become candidate materials for applications of spintronic and thermoelectric devices [5], [8], [9]. Recently, QHAs have received a good deal of interest due to their novel electronic, magnetic, and thermoelectric properties [10], [11], [12], [13]. In addition, the electronic devices depending on the QHAs are anticipated to have lower-power dissipation [14]. Although some QHAs such as NiFeMnGa and NiCoMnGa have shown excellent half-metallic behaviors, another such as CuCoMnGa exhibited a metallic behavior [15]. Using the plane-wave pseudopotential method, Li *et al.* predicted the half-metallicity of Nb X′ CrAl (X′=Co, Rh) quaternary Heusler alloys [16]. They also found that NbRhCrAl alloy became more robust against the lattice thermal expansion as the temperature increase [16]. Haleoot and Hamad have performed theoretical investigation of CoFeCuZ (Z = Al, As, Ga, In, Pb, Sb, Si, Sn) QHAs [4]. They found that CoFeCuPb alloy exhibits a half-metallic ferromagnetic structure with a spin-minority band gap of 0.303 eV and a total magnetic moment of 4.00 $\mu_B$ [4]. However, the other alloys were found to be either metallic for Z = (Al, As) or nearly half-metallic for Z = (Ga, In, Sb, Si, Sn) [4]. Bainsla *et al.* have reported that the CoFeMnGe QHA has a cubic structure of Y-type, and there is no phase transition at higher temperatures up to the melting temperature (1400 K) [17]. Moreover, ZrFeVZ (Z = Al, Ga, and In) QHAs were found to be excellent half-metallic ferromagnets with band gaps of 0.348, 0.428, and 0.323eV, respectively [18]. In addition to the magnetic structure of QHAs, several investigations predicted very interesting thermoelectric properties such as tunable lattice thermal conductivity, large Seebeck coefficient, and high thermoelectric functioning [11], [19], [20], [21]. The thermoelectric efficiency of these QHAs can be inferred by the dimensionless figure of merit $\left(ZT = \frac{S^2 \sigma T}{\kappa_e + \kappa_L}\right)$, where $S$ is the Seebeck coefficient, $\sigma$ is the electrical conductivity, $T$ is absolute temperature and $\kappa_e$ and $\kappa_L$ refer to the electronic and lattice thermal conductivities, respectively [22], [23]. The optimal thermoelectric material (TE) should have large values of $S$, and $\sigma$, that lead to the large power factor PF= $S^2 \sigma$, and small values of $\kappa_e$ and $\kappa_L$. Recently, Mushtaq *et al.* [24] have performed a theoretical study of two new semiconducting QHAs CoCuZrGe and CoCuZrSn by using FP-LAPW technique. They predicted Seebeck coefficient values of 26.2 µV/K and 28 µV/K for CoCuZrGe and CoCuZrSn alloys, respectively with a *p*-type semiconducting behavior. Furthermore, the PF of CoCuZrGe and CoCuZrSn were found to be $1.55 \times 10^{12}$ WK$^{-2}$m$^{-1}$s$^{-1}$ and $1.38 \times 10^{12}$ WK$^{-2}$m$^{-1}$s$^{-1}$, respectively [24]. The investigations of these QHAs also contribute in expanding the database of HMMs with optimal spin polarization (100%) and good thermoelectric properties.

In this work, the VTiRhZ (Z = Al, Ga, In) QHAs are investigated using first-principal calculations. The investigations include the structural, dynamical, mechanical, electronic, magnetic, and thermoelectric properties of VTiRhZ QHAs. The paper is arranged as follows: section 2 presents the computational details, section 3 includes the results and discussion, and section 4 summarizes the main conclusions.

## 2. Computational Details

The calculations are based on density functional theory (DFT). The structural optimizations were performed using the projector augmented wave (PAW) method as implemented in Vienna ab initio simulation package (VASP) code [25]. In these calculations, the plane-waves were expanded up to a cut-off energy of 520 eV with a total energy tolerance of $10^{-8}$ eV. For the formation energy calculations, the Brillouin zone integration was established with $22 \times 22 \times 22$ k-point mesh for unit-cell structures. The phonon calculation are obtained using phonopy package as implemented in VASP code [26] with $4 \times 4 \times 4$ supercell structures. These optimized parameters are then used to perform total energy calculations based on full-potential linearized augmented plane wave (FP-LAPW) method as implemented in WIEN2k code [27]. The exchange-correlation potential is treated by using Perdew–Burke–Ernzerhof Generalized Gradient Approximation (GGA-PBE) [28]. The wavefunctions in the interstitial region were described by plane waves with a cut-off value $K_{max} \times R_{MT}$=8.5, where $R_{MT}$ is the smallest atomic muffin tin radius and $K_{max}$ is the largest $k$ vectors in plane wave expansion. The $R_{MT}$ are chosen to be 2.4, 2.2, 2.0, and 1.9 atomic units (a.u.) for V, Ti, Rh, and Z atoms, respectively. The maximum angular momentum ($l_{max}$) inside the muffin-tin spheres was set to be 10 and the Fourier expansion of the charge density ($G_{max}$) was truncated at 12 (Ryd)$^{-1}$. In the self-consistent calculations, the total energy and charge density convergence tolerances were set to $10^{-4}$ Ry and $10^{-4}$ eV, respectively and the force tolerance was set up equal to 1 mRy/a. u. The thermoelectric parameters including the Seebeck coefficient, electrical conductivities, electronic thermal conductivity and power factor were calculated using Boltzmann transport theory, as implemented in the BoltzTrap code [29]. These TE properties are based on DFT calculations with a high dense mesh of $5.0 \times 10^4$ k-points, which is equivalent to a $36 \times 36 \times 36$ k-mesh. The TE calculations were performed within the constant relaxation time approximation.

### 3. Results and Discussions:

This section presents the structural, dynamical, mechanical, electronic, magnetic, and thermoelectric properties of VTiRhZ (Z = Al, Ga, In) alloys.

### 3.1 Structural properties

The chemical formula of VTiRhZ (Z = Al, Ga, In) QHAs is XX′YZ with 1:1:1:1 stoichiometry, where X, X′, and Y are transition metals and Z is an *s-p* element. The QHAs possess a face-centered cubic LiMgPdSn (Y-type) crystal structure with a space group $F\overline{4}3m$ (*no.* 216). In this type, the QHAs have three possible atomic configurations identified as LiMgPdSn (Y-type) crystal structures, see Table 1 and Fig. 1. The ground state configuration of each QHAs is identified by the standard energy minimization techniques, where the type-1 structure was found to be the most preferred structure, see Table 2. These results are in agreement with those of similar alloys such as CoFeMnZ (Z=Al, Ga, Si, Ge) [14]. The thermodynamic stability of these alloys is examined by the formation energy using the following equation [4]:

$$E_{form} = E_{tot} - \left(E_X^{bulk} + E_{X'}^{bulk} + E_Y^{bulk} + E_Z^{bulk}\right), \qquad (1)$$

Here $E_{tot}$ refers the total energy of the QHAs per formula unit, whereas $E_X^{bulk}$, $E_{X'}^{bulk}$, $E_Y^{bulk}$ and $E_Z^{bulk}$ refer to the total energies per atom in the alloys. The formation energy values of these alloys are presented in Table 3. All energies are found to be negative, which indicates the thermodynamic stability of VTiRhZ (Z= Al, Ga, In) QHAs in their type-I configuration. The optimized lattice parameter for each alloy is presented in Table 3.

### 3.2 Dynamical properties

This subsection presents the phonon calculations and dispersions relations to provide a better understanding of the dynamic stability of the investigated systems. The phonon dispersion curves (PDCs) of these alloys are obtained using phonopy package as implemented in VASP code [26]. The PDCs are depicted along the high symmetry k-path (W→L→Γ→X→W) in the first Brillouin zone, see Fig. 2(a, b, c). These curves show only positive frequencies without any imaginary (negative) frequencies for the three alloys, which confirms their dynamic stability in the type-I configuration. The unit cell contains four atoms (N=4), which leads to twelve phonon branches (3N), three acoustic and nine optical branches at the lower and higher frequencies, respectively, see Fig. 2. The three acoustic branches are composed of one longitudinal (LA) and two transverse (TA) modes. This figure shows that VTiRhAl, VTiRhGa, VTiRhIn alloys exhibit no phonon band gaps between acoustic and optical branches. This property is advantageous for the high power factor and low thermal conductivity [30].

### 3.3 Mechanical properties

In this subsection, the elastic constants $C_{ij}$ are calculated to provide a critical information about mechanical properties of the VTiRhZ QHAs. The cubic structure possesses three independent elastic constants, namely, $C_{11}$, $C_{12}$, and $C_{44}$, which refer to the longitudinal compression, transverse expansion, and share modulus predictor, respectively. There are three conditions of the Born and Huang criteria that should be satisfied to indicate the mechanical stability of the cubic structure given as [31]:

$$C_{44} > 0, (C_{11} - C_{12})/2 > 0, (C_{11} + 2C_{12})/3 > 0, \qquad (2)$$

According to these criteria, VTiRhZ alloys are mechanically stability in the type-I structure, see Table 3. Moreover, other mechanical parameters such as the bulk modulus ($B$), Voigt-Reuss shear modulus ($G$), Young's modulus ($E$), anisotropy factor ($A$), Poisson's ratios ($v$), Pugh's ratio ($B/G$) and Cauchy pressure ($C_P$) can be investigated by using the independent elastic constants [32], [33], [34], [35]. The calculated values of these mechanical parameters are presented in Table 3.

The bulk modulus ($B$) measures the resistance of a material to compressions and is defined as follows:

$$B = \frac{(C_{11} + 2C_{12})}{3}. \qquad (3)$$

The lowest $B$ value of 152.1 GPa was predicted for the case of VTiRhIn alloy, due to its higher lattice parameter, than those of VTiRhAl 168.2GPa and VTiRhGa 171.4 alloys, see Table 3. The shear modulus ($G$) gives information about the change of the shape due to an applied force, which is defined as the average of Voigt's shear ($G_V$) and Reuss's shear ($G_R$) moduli as follows:

$$G = \frac{(G_V + G_R)}{2},$$ (4)

where:

$$G_V = \frac{C_{11} - C_{12} + 3C_{44}}{5},$$ (5)

$$G_R = \frac{(5C_{44}(C_{11} - C_{12}))}{4C_{44} + 3(C_{11} - C_{12})},$$ (6)

The largest $G$ value of 93.6 GPa was obtained for VTiRhAl alloy. In addition, Young's modulus ($E$) provides a measure of the stiffness of a material, which is expressed in terms of $G$ and $B$ as follows:

$$E = \frac{9GB}{3B + G}.$$ (7)

The highest $E$ value was obtained for the case of VTiRhAl alloy, which indicates that it is the stiffest as compared to VTiRhGa and VTiRhIn alloys, see Table 3.

Poisson's ratio (v) provides a measure of the compressibility of the material, which is expressed as follows:

$$v = \frac{3B - 2G}{2(3B + G)}.$$ (8)

Poisson's ratio usually ranges between 0.25 to 0.5 [36]. Materials with more than 0.26 are considered as the ductile, whereas those with values less than 0.26 [37] are brittle. In this work, the values of Poisson's ratio are 0.27, 0.28, and 0.29 for VTiRhAl, VTiRhGa and VTiRhIn, respectively, which indicate that these QHAs are stable and ductile. The results are comparable to previous calculations of similar structures such as CoCuZrSn (v = 0.38) and CoCuMnSn (v = 0.28) [37].

Another important quantity for the measure of stability is Cauchy pressure, which is defined as follows:

$$C_p = C_{12} - C_{44} \qquad (9)$$

The material is considered to be ductile if its Cauchy pressure value is positive, otherwise it is brittle [38]. The Cauchy pressure of VTiRhZ (Z=Al, Ga, In) alloys are found to be positive values, which proves that these alloys are ductile in nature. In addition, the *B/G* calculation also gives information about ductility and brittleness of materials. While ductile materials possess *B/G* > 1.75, brittle materials have *B/G* < 1.75 [39]. The values of *B/G* for VTiRhAl, VTiRhGa and VTiRhIn alloys are 1.76, 1.99 and 2.15, respectively, which further confirms the ductile nature of VTiRhZ alloys.

The elastic anisotropy factor is another important quantity, which is defined as follows:

$$A = \frac{2C_{44}}{C_{11} - C_{12}}. \qquad (10)$$

Isotropic materials have an anisotropy factor *A*=1, whereas, anisotropic materials exhibit *A* values more or less than unity [40]. Accordingly, VTiRhZ alloys are considered to be anisotropic since the anisotropy factor values are less than unity, see Table 3. This is consistent with previous calculations of the anisotropy factor for CoFeCrGe (0.62) and CoFeTiGe(0.76) QHAs[13].

The melting temperature of materials identifies the heat resistance of the material, which can be computed with following equation [38], [41]:

$$T_{melt} = \left[ 553\text{K} + \left( \frac{5.91\text{K}}{GPa} \right) C_{11} \right] \pm 300\text{K} \qquad (11)$$

From this equation, the melting temperature values of VTiRhAl, VTiRhGa, and VTiRhIn are found to be 2312 K, 2268 K and 2028 K, respectively. These high melting temperatures indicate the stability of these alloys within an error of $\pm 300$K. The melting temperature was found to decrease by increasing the atomic number of Z atoms.

### 3.4 Electronic properties

This subsection introduces the electronic structure of VTiRhZ (Z= Al, Ga, In) quaternary Heusler alloys. Figure 3(a, b, c) presents the band structures and total density of states (TDOS) of VTiRhZ (Z= Al, Ga, In) alloys in their stable configuration (type-I) along the high symmetry k-path. This figure shows that VTiRhAl alloy has a semiconducting behavior in both majority and minority spin channels with band gap values (0.04 and 0.62 eV), respectively. Both VTiRhGa and VTiRhIn possess a half metallic behavior (metallic majority spin channel and a semiconducting minority spin channel). The minority spin band gaps of VTiRhGa and VTiRhIn alloys are found to be indirect from the conduction band maximum (CBM) at the L high symmetry point to the valence band minimum (VBM) at the $\Gamma$ high symmetry point with 0.52 and 0.19 eV, respectively. It is obvious that the band gap in the minority spin channel decreases by increasing the atomic number of the Z atom (Al, Ga, and In), see Table 4. Moreover, the flat energy levels are presented in the $\Gamma$ -X symmetry line of the conduction bands. In addition, there is the extremely dispersive bands in other directions. These two properties could be a feature to increase the Seebeck coefficient value and power factor [42]. From Fig. 3 (a), one can that the valence band in the majority channel is exactly located at the Fermi level, which indicates that this alloy is classified as spin-gapless semiconductor materials. However, VTiRhGa and VTiRhIn alloys show a shift in of the valence bands above the Fermi level at the $\Gamma$ high symmetry point. This indicates the half-metallic behavior of these two alloys, which are very close to be spin-gapless semiconductor materials. These results are in agreement with previous ab initio investigations of (PtVScAl, PtVYAl, and PtVYGa) QHAs [43].

The projected density of states (PDOS) is presented in Figs. 4 (a, b, c). from this figure, the valence band of these alloys have two main regions. The first region between -4 eV to -2 eV is exhibited the main contribution of Rh-*d* orbital and small contribution of *d*- and *p*- orbitals of Ti and Z (Z= Al, Ga, In) atoms in both the majority and minority spin channels. However, the second region is between -2 eV to the Fermi level at zero point. This region shows a mixture of different orbitals in majority spin channel, whereas the minority spin channel is mainly contributed by *d*-orbital of V atom. In the conduction band for these alloys, the most significant contribution comes from atoms of V, Ti, and Rh-d orbital.

The spin polarization at the Fermi level is a key quantity that measures the half metallicity of structures, which can be calculated using the following equation [44]:

$$P = \frac{\rho_{\text{majority}}(E_f) - \rho_{\text{minority}}(E_f)}{\rho_{\text{majority}}(E_f) + \rho_{\text{minority}}(E_f)} \times 100 \qquad (12)$$

where $\rho_{\text{majority}}(E_f)$ and $\rho_{\text{minority}}(E_f)$ correspond to the majority and minority spin density of states at the Fermi level $E_f$, respectively [44]. The spin polarization value of 100% is a perfect half-metallicity due to a zero density of states at the Fermi level $E_f$ in either majority or minority spin channel. The spin polarization values of the alloys are listed in Table 4. A perfect spin polarization of 100% was obtained for both VTiRhGa and VTiRhIn alloys, which corresponds to a half-metallic behavior.

### 3.5 Magnetic properties

In this subsection, the magnetic properties are calculated for the dynamically stable VTiRhZ (Z= Al, Ga, In) QHAs. In general, half-metallic materials have integer values of magnetic moments (M) based on the slater-Pauling equation[45], [46]:

$$M_{tot} = N_{majority} - N_{minority} = (Z_{tot} - N_{minority}) - N_{minority} = Z_{tot} - 2N_{minority} \qquad (13)$$

where, $M_{tot}$, $N_{\text{majority}}$, $N_{\text{minority}}$, and $Z_{\text{tot}}$ are the total magnetic moment, the majority spin valence electrons, the minority spin valence electrons, and the total valence electron number, respectively. For the cases of VTiRhAl, VTiRhGa, and VTiRhIn, the total magnetic moment was found to be 3μ$_B$. These values can be calculated using the Slater-Pauling rule using the valence electron configurations of V (3d$^3$4s$^2$), Ti (3d$^2$4s$^2$), Rh (4d$^8$5s$^1$) and Z = (Al(3s$^2$3p$^1$), Ga (4s$^2$4p$^1$), and In (5s$^2$5p$^1$). Therefore, the total valence electron number for VTiRhAl, VTiRhGa, and VTiRhIn is Z$_{tot}$ = 21. Using the TDOS at the $E_f$, the valence electrons have 12 and 9 in the majority and minority spin channels, respectively. Thus, the total magnetic moment of VTiRhAl QHA is M$_{tot}$ = N$_{majority}$ − N$_{minority}$ = 12 − 9 = 3$\mu_B$, which is satisfied the Slater-Pauling rule M$_{tot}$ = N$_{majority}$ − N$_{minority}$ = (Z$_{tot}$ − N$_{minority}$) − N$_{minority}$ = Z$_{tot}$ − 2N$_{minority}$ = Z$_{tot}$ − 18. The local magnetic moments of V, Ti, Rh in the cases of VTiRhAl, VTiRhGa, VTiRhIn alloys are ferromagnetically coupled, where the V atoms show the highest magnetic moments of 2.19, 2.21, and 2.25 $\mu_B$, respectively. The values of the total and local magnetic moments per atom are presented in Table 4.

The linear relation between Curie temperature ($T_C$) and total magnetic moments is considered to be one of the methods that has been adopted to estimate the Curie temperature by using the following equation [4], [38], [47], [48]:

$$T_C = 23 + 181 \, M_{tot} \tag{19}$$

The value of Curie temperature for VTiRhZ alloys is found to be 566 K, which is higher than room temperature. Thus, these QHAs are appropriate for spintronics applications.

### 3.6 Thermoelectric properties

This subsection presents the thermoelectric properties of VTiRhZ quaternary Heusler alloys. The Boltzmann transport theory is applied to calculate the transport properties of the Seebeck coefficient ($S$), electrical conductivity ($\sigma/\tau$), and power factor per relaxation time ($S^2\sigma/\tau$). The solution of the Boltzmann transport equation has the following form [49]:

$$\frac{\partial f_{\vec{k}}}{\partial t} = -\vec{v_{\vec{k}}} \cdot \frac{\partial f_{\vec{k}}}{\partial \vec{r}} - \frac{e}{\hbar}\left(\vec{E} + \frac{1}{c}\vec{v_{\vec{k}}} \times \vec{H}\right) \cdot \frac{\partial f_{\vec{k}}}{\partial \vec{k}} + \frac{\partial f_{\vec{k}}}{\partial t}\big|_{scatt} \tag{14}$$

where $\vec{k}$ and $\overrightarrow{v_{\vec{k}}}$ are the wave vector and the group velocity, respectively. $f_{\vec{k}}$ expresses the occupation of the quantum state (the distribution function). Based on this solution, $f_{\vec{k}}$ is based on the applied electric ($\vec{E}$) and magnetic ($\vec{H}$) fields. The Seebeck coefficient and electrical conductivity are given as:

$$S_{\alpha\beta}(T,\mu) = \frac{1}{eT\Omega\sigma_{\alpha\beta}(T,\mu)} \int \bar{\sigma}_{\alpha\beta}(\varepsilon)(\varepsilon - \mu)\left[-\frac{\partial f_0(T,\varepsilon,\mu)}{\partial\varepsilon}\right]d\varepsilon \qquad (15)$$

$$\sigma_{\alpha\beta}(T,\mu) = \frac{1}{\Omega} \int \bar{\sigma}_{\alpha\beta}(\varepsilon)\left[-\frac{\partial f_0(T,\varepsilon,\mu)}{\partial\varepsilon}\right]d\varepsilon \qquad (16)$$

where α and β are tensor indices and μ, Ω, and $f_0$ are the chemical potential, unit cell volume and the Fermi-Dirac distribution function, respectively [41].

The thermoelectric properties of VTiRhZ alloys are investigated for the minority spin channel only. This is attributed to the fact that this channel exhibits a semiconducting behavior with a narrow band gap in addition to flat energy states along the Γ -X symmetry line in the conduction band. Narrow band gap semiconductors are believed to exhibit promising thermoelectric properties [50], [51], [52]. The Seebeck coefficient is plotted as a function of the chemical potential at 300 K and 800 K, see Figs. 5 (a) and (b). The Seebeck coefficient gives maximum values near the Fermi level, which decrease as the temperatures increases. At each temperature, the highest and lowest values of the Seebeck coefficient are obtained for VTiRhAl and VTiRhIn alloys, respectively. In The electrical conductivity per relaxation time ($\sigma/\tau$) is plotted as a function of the chemical potential in Figs. 5 (c) and 5 (d) at 300K and 800K, respectively, which vanishes around the Fermi-level as a typical behavior of semiconductors. In addition, the *n*-type doping reveals higher values of $\sigma/\tau$ than the *p*-type for the investigated QHAs. The values of the power factor as a function of the chemical potential at 300 K and 800 K are also presented in Fig. 5 (e) and (f). Unlike the Seebeck coefficient, the *PF* values are found to be higher at 800K than 300K. From this figure, one can notice that the highest and lowest *PF* values at 800 K are $8.2 \times 10^{11}$ W.m$^{-1}$ K$^{-2}$ s$^{-1}$ and $14 \times 10^{11}$ W.m$^{-1}$ K$^{-2}$ s$^{-1}$ for VTiRhAl and VTiRhIn QHAs, respectively. These predicted values are higher than that obtained in a previous study for CoNbMnAl quaternary Heusler alloy of $6.9 \times 10^{11}$ Wm$^{-1}$ K$^{-2}$ s$^{-1}$ [46].

The lattice thermal conductivity ($\kappa_l$) is an essential quantity to obtain the figure of merit (ZT) for these investigated QHAs. Slack's formula is considered to be one of the methods that has been successfully used for similar materials[53], [54], [55], [56] to calculate the $\kappa_l$ value using the following equation [57]:

$$\kappa_l = A \frac{\bar{M} \Theta_D^3 V^{1/3}}{\gamma^2 n^{2/3} T} \tag{17}$$

Here *A* is constant that is given by $\left( \frac{2.43 \times 10^{-6}}{1 - \frac{0.514}{\gamma} + \frac{0.228}{\gamma^2}} \right)$ [57], and $\bar{M}$, $\Theta_D$, V, $\gamma$, *n* and *T* refer to the average atomic mass, Debye temperature, volume per atom, Grüneisen parameter, number of atoms in the primitive unit cell, and temperature, respectively. Based on the elastic constant calculations, the Debye temperature and Grüneisen parameter are calculated by using the following equations [57]:

$$\Theta_D = \frac{h}{k_B} \left( \frac{3n}{4\pi\Omega} \right)^{1/3} v_m \quad , \tag{18}$$

$$v_m = \left[ \frac{1}{3} \left( \frac{2}{v_t^3} + \frac{1}{v_l^3} \right) \right]^{-1/3} , \tag{19}$$

$$v_l = \sqrt{\frac{3B + 4G}{3\rho}} \quad , \tag{20}$$

$$v_t = \sqrt{\frac{G}{\rho}} \quad , \tag{21}$$

$$\gamma = \frac{9-12(v_t/v_l)^2}{2+4(v_t/v_l)^2} \ , \qquad\qquad\qquad\qquad\qquad\qquad\qquad (22)$$

The parameters $h$, $k_B$, $\Omega$, $\rho$, $v_m$, $v_l$ and $v_t$ refer to the Planck constant, Boltzmann constant, cell volume, density, average, transverse, and longitudinal sound velocities, respectively. The Debye temperature $\Theta_D$, average, transverse, longitudinal sound velocities, density and Grüneisen parameter $\gamma$ are listed in Table 5. The Debye temperature of VTiRhAl (514.11K) is found to be higher than those of VTiRhGa (447.5K) and VTiRhIn (382.8K), which indicates that the lattice thermal conductivity is higher for VTiRhAl than those of VTiRhGa and VTiRhIn alloys. From this table, one can notice that the Debye temperature decreases by increasing the atomic number of the Z atom (Al, Ga, In). These results are in agreement with other previous calculations of Co$_2$MnZ (Z=Al, Ga, In) [41]. The lattice thermal conductivity was calculated using the aforementioned parameters, see Fig. 6. From this figure, it is obvious that VTiRhAl has higher values of lattice thermal conductivity than those of VTiRhGa and VTiRhIn alloys. The electronic thermal conductivity $\kappa_e$ and figure of merit $ZT$ values are plotted as a function of the chemical potential at 300 K and 800 K, see Fig. 7. The electronic thermal conductivities of the investigated QHAs were found to be higher at 800K than 300K. The $n$-type exhibits higher $\kappa_e$ values than $p$-type in both temperatures, see Fig. 7 (a) and (b). The $ZT$ values of VTiRhZ (Z=Al, Ga, In) are found to be higher at 300K than 800K, which is opposite to the behavior of the lattice thermal and electronic conductivities (Fig.6 and Fig.7 (a) and (b)). There are two peaks of $ZT$ values for each system. At 300K, the VTiRhAl and VTiRhGa exhibit $p$-type behavior with $ZT$ values of 0.96, and 0.88, respectively, while VTiRhIn shows both $p$- and $n$-type behaviors with $ZT$ values of 0.54 and 0.64, respectively, see Fig. 7 (c). However, all structures exhibit both $p$- and $n$-type behaviors at 800K. At this temperature, the two peaks of VTiRhAl show the highest $ZT$ value of 0.85 and 0.69 in $p$-type and $n$-type, respectively, see Fig. 7 (d). These results of $ZT$ values are higher than other similar previous calculations of 0.65 eV and 0.71 eV for CoFeTiGe and CoFeCrGe QHAs, respectively [13]. Therefore, VTiRhZ (Z=Al, Ga, In) QHAs are good candidates for further theoretical and experimental investigations in low dimensional and doped systems that may provide higher *ZT* values for promising TE applications.

## 4. Conclusion

The structural, dynamical, mechanical, electronic, magnetic, and thermoelectric properties of VTiRhZ (Z = Al, Ga, In) alloys are investigated using DFT calculations. These alloys are found to be stable in the type-I structure. The GGA-PBE calculations predict a half metallic ferromagnetic behavior for VTiRhZ (Z =Ga, In) alloys with band gaps of 0.52, and 0.19, respectively. However, VTiRhAl shows a ferromagnetic behavior with a semiconducting structure in both spin channels. The VTiRhZ (Z= Ga, In) alloys possess a total magnetic moment of $3\mu_B$ and a spin polarization of 100%, which suggest them as prominent candidates for spin-injection. Using the semi-classical Boltzmann transport theory within the constant relaxation time approximation, VTiRhZ (Z=Al, Ga, In) alloys show good thermoelectric properties. The highest value of the power factor per relaxation time is $14 \times 10^{11}$ Wm$^{-1}$ K$^{-2}$ s$^{-1}$ for VTiRhAl. In addition, VTiRhAl, VTiRhGa, and VTiRhIn alloys show high figure of merit values of 0.96, 0.88 and 0.64, respectively at the room temperature. Thus, these alloys can find significant applications as thermoelectric materials at moderate temperatures.

## Acknowledgement

Hind Alqurashi was financially supported by Al-Baha university and Saudi Arabian Cultural Mission. The calculations were performed on the high-performance computing of the University of Arkansas.

**Conflict of interest**

The authors declare that they have no conflict of interest.

**References:**

[1]     T. Graf, F. Casper, J. Winterlik, B. Balke, G.H. Fecher, C. Felser, Crystal Structure of New Heusler Compounds, Zeitschrift Für Anorg. Und Allg. Chemie. 635 (2009) 976–981. https://doi.org/10.1002/zaac.200900036.

[2]     G. Qin, W. Wu, S. Hu, Y. Tao, X. Yan, C. Jing, X. Li, H. Gu, S. Cao, W. Ren, Effect of swap disorder on the physical properties of the quaternary Heusler alloy PdMnTiAl: A first-principles study, IUCrJ. 4 (2017) 506–511. https://doi.org/10.1107/S205225251700745X.

[3]     S. Skaftouros, K. Özdoğan, E. Şaşioğlu, I. Galanakis, Generalized Slater-Pauling rule for the inverse Heusler compounds, Phys. Rev. B - Condens. Matter Mater. Phys. 87 (2013) 024420. https://doi.org/10.1103/PhysRevB.87.024420.

[4]     R. Haleoot, B. Hamad, Ab Initio Investigations of the Structural, Electronic, Magnetic, and Thermoelectric Properties of CoFeCuZ (Z = Al, As, Ga, In, Pb, Sb, Si, Sn) Quaternary Heusler Alloys, J. Electron. Mater. 48 (2019) 1164–1173. https://doi.org/10.1007/s11664-018-6833-1.

[5]     L. Bainsla, K.G. Suresh, Equiatomic quaternary Heusler alloys: A material perspective for spintronic applications, Appl. Phys. Rev. 3 (2016) 031101. https://doi.org/10.1063/1.4959093.

[6]     I. Galanakis, P. Mavropoulos, P.H. Dederichs, Electronic structure and Slater-Pauling behaviour in half-metallic Heusler alloys calculated from first principles, J. Phys. D. Appl. Phys. 39 (2006) 765–775. https://doi.org/10.1088/0022-3727/39/5/S01.

[7]     R.A. De Groot, F.M. Mueller, P.G.V. Engen, K.H.J. Buschow, New class of materials: Half-metallic ferromagnets, Phys. Rev. Lett. 50 (1983) 2024–2027.

https://doi.org/10.1103/PhysRevLett.50.2024.

[8]    B. Hamad, Theoretical Investigations of the Thermoelectric Properties of

Fe2NbGa1−xAlx (x = 0, 0.25, 0.5) Alloys, J. Electron. Mater. 46 (2017) 6595–6602.

https://doi.org/10.1007/s11664-017-5721-4.

[9]    G. Joshi, B. Poudel, Efficient and Robust Thermoelectric Power Generation Device Using

Hot-Pressed Metal Contacts on Nanostructured Half-Heusler Alloys, J. Electron. Mater.

45 (2016) 6047–6051. https://doi.org/10.1007/s11664-016-4692-1.

[10]   X. Guo, Z. Ni, Z. Liang, H. Luo, Magnetic semiconductors and half-metals in FeRu-based

quaternary Heusler alloys, Comput. Mater. Sci. 154 (2018) 442–448.

https://doi.org/10.1016/j.commatsci.2018.08.023.

[11]   L. Zhang, Z.X. Cheng, X.T. Wang, R. Khenata, H. Rozale, First-Principles Investigation

of Equiatomic Quaternary Heusler Alloys NbVMnAl and NbFeCrAl and a Discussion of the Generalized Electron-Filling Rule, J. Supercond. Nov. Magn. 31 (2018) 189–196.

https://doi.org/10.1007/s10948-017-4182-6.

[12]   C.K. Barman, C. Mondal, B. Pathak, A. Alam, Quaternary Heusler alloy: An ideal

platform to realize triple point fermions, Phys. Rev. B. 99 (2019) 045144.

https://doi.org/10.1103/PhysRevB.99.045144.

[13]   R. Haleoot, B. Hamad, Thermodynamic and thermoelectric properties of CoFeYGe (Y =

Ti, Cr) quaternary Heusler alloys: First principle calculations, J. Phys. Condens. Matter.

32 (2020) 075402. https://doi.org/10.1088/1361-648X/ab5321.

[14]   V. Alijani, S. Ouardi, G.H. Fecher, J. Winterlik, S.S. Naghavi, X. Kozina, G. Stryganyuk,

C. Felser, E. Ikenaga, Y. Yamashita, S. Ueda, K. Kobayashi, Electronic, structural, and

magnetic properties of the half-metallic ferromagnetic quaternary Heusler compounds

CoFeMnZ (Z=Al, Ga, Si, Ge), Phys. Rev. B - Condens. Matter Mater. Phys. 84 (2011) 224416. https://doi.org/10.1103/PhysRevB.84.224416.

[15]   V. Alijani, J. Winterlik, G.H. Fecher, S.S. Naghavi, C. Felser, Quaternary half-metallic Heusler ferromagnets for spintronics applications, Phys. Rev. B - Condens. Matter Mater. Phys. 83 (2011) 184428. https://doi.org/10.1103/PhysRevB.83.184428.

[16]   Y. Li, G.D. Liu, X.T. Wang, E.K. Liu, X.K. Xi, W.H. Wang, G.H. Wu, L.Y. Wang, X.F. Dai, First-principles study on electronic structure, magnetism and half-metallicity of the NbCoCrAl and NbRhCrAl compounds, Results Phys. 7 (2017) 2248–2254. https://doi.org/10.1016/j.rinp.2017.06.047.

[17]   L. Bainsla, K.G. Suresh, A.K. Nigam, M.M. Raja, B.S.D.C.S. Varaprasad, Y.K. Takahashi, K. Hono, High spin polarization in CoFeMnGe equiatomic quaternary Heusler alloy, J. Appl. Phys. 116 (2014) 203902. https://doi.org/10.1063/1.4902831.

[18]   R. Guo, G. Liu, X. Wang, H. Rozale, L. Wang, R. Khenata, Z. Wu, X. Dai, First-principles study on quaternary Heusler compounds ZrFeVZ (Z = Al, Ga, In) with large spin-flip gap, RSC Adv. 6 (2016) 109394–109400. https://doi.org/10.1039/c6ra18873g.

[19]   S.N.H. Eliassen, A. Katre, G.K.H. Madsen, C. Persson, O.M. Løvvik, K. Berland, Lattice thermal conductivity of TixZryHf1-x-yNiSn half-Heusler alloys calculated from first principles: Key role of nature of phonon modes, Phys. Rev. B. 95 (2017) 045202. https://doi.org/10.1103/PhysRevB.95.045202.

[20]   C.S. Lue, C.F. Chen, J.Y. Lin, Y.T. Yu, Y.K. Kuo, Thermoelectric properties of quaternary Heusler alloys Fe2 VAl1-x Six, Phys. Rev. B - Condens. Matter Mater. Phys. 75 (2007) 064204. https://doi.org/10.1103/PhysRevB.75.064204.

[21]   S. Yabuuchi, M. Okamoto, A. Nishide, Y. Kurosaki, J. Hayakawa, Large seebeck

coefficients of fe2tisn and fe2tisi: First-principles study, Appl. Phys. Express. 6 (2013) 025504. https://doi.org/10.7567/APEX.6.025504.

[22]   T.T. Lin, Q. Gao, G.D. Liu, X.F. Dai, X.M. Zhang, H.B. Zhang, Dynamical stability, electronic and thermoelectric properties of quaternary ZnFeTiSi Heusler compound, Curr. Appl. Phys. 19 (2019) 721–727. https://doi.org/10.1016/J.CAP.2019.03.020.

[23]   H. Alqurashi, R. Haleoot, A. Pandit, B. Hamad, Investigations of the electronic, dynamical, and thermoelectric properties of Cd1-xZnxO alloys: First-principles calculations, Mater. Today Commun. 28 (2021) 102511. https://doi.org/10.1016/j.mtcomm.2021.102511.

[24]   M. Mushtaq, M.A. Sattar, S.A. Dar, Phonon phase stability, structural, mechanical, electronic, and thermoelectric properties of two new semiconducting quaternary Heusler alloys <scp>CoCuZrZ</scp> (Z = Ge and Sn), Int. J. Energy Res. 44 (2020) 5936–5946. https://doi.org/10.1002/er.5373.

[25]   D. Joubert, G. Kresse, D. Joubert, From Ultrasoft Pseudopotentials to the Projector Augmented-Wave Method First-Principles Calculations of the Electronic and Optical Properties of CH 3 NH 3 PbI 3 for Photovoltaic Applications View project One dimensional Lattice Density Functional Theory View project From ultrasoft pseudopotentials to the projector augmented-wave method, Artic. Phys. Rev. B. (1999). https://doi.org/10.1103/PhysRevB.59.1758.

[26]   G. Kresse, J. Hafner, Ab initio molecular dynamics for liquid metals, Phys. Rev. B. 47 (1993) 558–561. https://doi.org/10.1103/PhysRevB.47.558.

[27]   P. Blaha, K. Schwarz, P. Sorantin, S.B. Trickey, Full-potential, linearized augmented plane wave programs for crystalline systems, Comput. Phys. Commun. 59 (1990) 399–

415. https://doi.org/10.1016/0010-4655(90)90187-6.

[28] J.P. Perdew, K. Burke, M. Ernzerhof, Generalized gradient approximation made simple, Phys. Rev. Lett. 77 (1996) 3865–3868. https://doi.org/10.1103/PhysRevLett.77.3865.

[29] G.K.H. Madsen, D.J. Singh, BoltzTraP. A code for calculating band-structure dependent quantities, Comput. Phys. Commun. 175 (2006) 67–71. https://doi.org/10.1016/j.cpc.2006.03.007.

[30] D.M. Hoat, M. Naseri, Electronic and thermoelectric properties of RbYSn half-Heusler compound with 8 valence electrons: Spin-orbit coupling effect, Chem. Phys. 528 (2020) 110510. https://doi.org/10.1016/j.chemphys.2019.110510.

[31] Dynamical theory of crystal lattices - CERN Document Server, (n.d.). https://cds.cern.ch/record/224197 (accessed April 21, 2021).

[32] R. Hill, The elastic behaviour of a crystalline aggregate, Proc. Phys. Soc. Sect. A. 65 (1952) 349–354. https://doi.org/10.1088/0370-1298/65/5/307.

[33] J.S. Zhao, Q. Gao, L. Li, H.H. Xie, X.R. Hu, C.L. Xu, J.B. Deng, First-principles study of the structure, electronic, magnetic and elastic properties of half-Heusler compounds LiXGe (X = Ca, Sr and Ba), Intermetallics. 89 (2017) 65–73. https://doi.org/10.1016/j.intermet.2017.04.011.

[34] X. Wang, Z. Cheng, J. Wang, G. Liu, A full spectrum of spintronic properties demonstrated by a C1b-type Heusler compound Mn2Sn subjected to strain engineering, J. Mater. Chem. C. 4 (2016) 8535–8544. https://doi.org/10.1039/c6tc02526a.

[35] F. Semari, R. Boulechfar, F. Dahmane, A. Abdiche, R. Ahmed, S.H. Naqib, A. Bouhemadou, R. Khenata, X.T. Wang, Phase stability, mechanical, electronic and thermodynamic properties of the Ga3Sc compound: An ab-initio study, Inorg. Chem.

Commun. 122 (2020) 108304. https://doi.org/10.1016/j.inoche.2020.108304.

[36] A. Candan, Magnetic, Electronic, Mechanic, Anisotropic Elastic and Vibrational Properties of Antiferromagnetic Ru2TGa (T = Cr, Mn, and Co) Heusler Alloys, J. Electron. Mater. 48 (2019) 7608–7622. https://doi.org/10.1007/s11664-019-07625-5.

[37] M. Mushtaq, M.A. Sattar, S.A. Dar, Phonon phase stability, structural, mechanical, electronic, and thermoelectric properties of two new semiconducting quaternary Heusler alloys <scp>CoCuZrZ</scp> (Z = Ge and Sn), Int. J. Energy Res. 44 (2020) 5936–5946. https://doi.org/10.1002/er.5373.

[38] R. Jain, V.K. Jain, A.R. Chandra, V. Jain, N. Lakshmi, Study of the Electronic Structure, Magnetic and Elastic Properties and Half-Metallic Stability on Variation of Lattice Constants for CoFeCrZ (Z = P, As, Sb) Heusler Alloys, J. Supercond. Nov. Magn. 31 (2018) 2399–2409. https://doi.org/10.1007/s10948-017-4460-3.

[39] X.-R. Chen, M.-M. Zhong, Y. Feng, Y. Zhou, H.-K. Yuan, H. Chen, Structural, electronic, elastic, and thermodynamic properties of the spin-gapless semiconducting Mn $_2$ CoAl inverse Heusler alloy under pressure, Phys. Status Solidi. 252 (2015) 2830–2839. https://doi.org/10.1002/pssb.201552389.

[40] Y. Lv, X. Zhang, W. Jiang, Phase stability, elastic, anisotropic properties, lattice dynamical and thermodynamic properties of B12M (M=Th, U, Np, Pu) dodecaborides, Ceram. Int. 44 (2018) 128–135. https://doi.org/10.1016/j.ceramint.2017.09.147.

[41] S.C. Wu, G.H. Fecher, S. Shahab Naghavi, C. Felser, Elastic properties and stability of Heusler compounds: Cubic Co $_2$ YZ compounds with L $2_1$ structure, J. Appl. Phys. 125 (2019) 082523. https://doi.org/10.1063/1.5054398.

[42] E.M. Levin, Charge carrier effective mass and concentration derived from combination of

Seebeck coefficient and Te 125 NMR measurements in complex tellurides, Phys. Rev. B. 93 (2016) 245202. https://doi.org/10.1103/PhysRevB.93.245202.

[43] Q. Gao, I. Opahle, H. Zhang, High-throughput screening for spin-gapless semiconductors in quaternary Heusler compounds, Phys. Rev. Mater. 3 (2019) 024410. https://doi.org/10.1103/PhysRevMaterials.3.024410.

[44] M. Beth Stearns, Simple explanation of tunneling spin-polarization of Fe, Co, Ni and its alloys, J. Magn. Magn. Mater. 5 (1977) 167–171. https://doi.org/10.1016/0304-8853(77)90185-8.

[45] I. Galanakis, P.H. Dederichs, N. Papanikolaou, Slater-Pauling behavior and origin of the half-metallicity of the full-Heusler alloys, Phys. Rev. B - Condens. Matter Mater. Phys. 66 (2002) 1–9. https://doi.org/10.1103/PhysRevB.66.174429.

[46] A.A. Mubarak, S. Saad, F. Hamioud, M. Al-Elaimi, Structural, thermo-elastic, electro-magnetic and thermoelectric attributes of quaternary CoNbMnX (X = Al, Si) Heusler alloys, Solid State Sci. 111 (2021) 106397. https://doi.org/10.1016/j.solidstatesciences.2020.106397.

[47] A. Candan, G. Uğur, Z. Charifi, H. Baaziz, M.R. Ellialtioğlu, Electronic structure and vibrational properties in cobalt-based full-Heusler compounds: A first principle study of Co2MnX (X = Si, Ge, Al, Ga), J. Alloys Compd. 560 (2013) 215–222. https://doi.org/10.1016/j.jallcom.2013.01.102.

[48] M.H. Elahmar, H. Rached, D. Rached, R. Khenata, G. Murtaza, S. Bin Omran, W.K. Ahmed, Structural, mechanical, electronic and magnetic properties of a new series of quaternary Heusler alloys CoFeMnZ (Z=Si, As, Sb): A first-principle study, J. Magn. Magn. Mater. 393 (2015) 165–174. https://doi.org/10.1016/j.jmmm.2015.05.019.

[49] J. Scheidemantel, C. Ambrosch-Draxl, T. Thonhauser, V. Badding, O. Sofo, Transport coefficients from first-principles calculations, Phys. Rev. B - Condens. Matter Mater. Phys. 68 (2003) 125210. https://doi.org/10.1103/PhysRevB.68.125210.

[50] J.O. Sofo, G.D. Mahan, Optimum band gap of a thermoelectric material, Phys. Rev. B. 49 (1994) 4565–4570. https://doi.org/10.1103/PhysRevB.49.4565.

[51] Z.M. Gibbs, H.S. Kim, H. Wang, G.J. Snyder, Band gap estimation from temperature dependent Seebeck measurement - Deviations from the 2e|S|maxTmax relation, Appl. Phys. Lett. 106 (2015) 022112. https://doi.org/10.1063/1.4905922.

[52] E.H. Hasdeo, L.P.A. Krisna, M.Y. Hanna, B.E. Gunara, N.T. Hung, A.R.T. Nugraha, Optimal band gap for improved thermoelectric performance of two-dimensional Dirac materials, J. Appl. Phys. 126 (2019) 035109. https://doi.org/10.1063/1.5100985.

[53] A.J. Hong, L. Li, R. He, J.J. Gong, Z.B. Yan, K.F. Wang, J.M. Liu, Z.F. Ren, Full-scale computation for all the thermoelectric property parameters of half-Heusler compounds, Sci. Rep. 6 (2016) 1–12. https://doi.org/10.1038/srep22778.

[54] H. Ma, C.L. Yang, M.S. Wang, X.G. Ma, Y.G. Yi, Effect of M elements (M = Ti, Zr, and Hf) on thermoelectric performance of the half-Heusler compounds MCoBi, J. Phys. D. Appl. Phys. 52 (2019) 255501. https://doi.org/10.1088/1361-6463/ab137d.

[55] C. Dhakal, S. Aryal, R. Sakidja, W.Y. Ching, Approximate lattice thermal conductivity of MAX phases at high temperature, J. Eur. Ceram. Soc. 35 (2015) 3203–3212. https://doi.org/10.1016/j.jeurceramsoc.2015.04.013.

[56] R. Haleoot, B. Hamad, Thermoelectric properties of doped β-InSe by Bi: First principle calculations, Phys. B Condens. Matter. 587 (2020) 412105. https://doi.org/10.1016/j.physb.2020.412105.

[57] G.A. Slack, Nonmetallic crystals with high thermal conductivity, J. Phys. Chem. Solids. 34 (1973) 321–335. https://doi.org/10.1016/0022-3697(73)90092-9.

**List of Figures:**

**Figure. 1:** The conventional cells of VTiRhZ (Z=Al, Ga, In) quaternary Heusler alloys in the three types of configurations.

**Figure. 2:** The phonon dispersion curves of (a) VTiRhAl, (b) VTiRhGa, (c) VTiRhIn quaternary Heusler alloys.

**Figure. 3:** The electronic band structures and total density of states (TDOS) of (a) VTiRhAl, (b) VTiRhGa, (c) VTiRhIn. The solid and dotted lines represent the majority and minority spin channels, respectively.

**Figure. 4:** The projected density of state (PDOS) of (a) VTiRhAl, (b) VTiRhGa, (c) VTiRhIn for the majority and minority spin channels.

**Figure. 5:** (a) and (b) the Seebeck coefficient (S), (c) and (d) electrical conductivity per relaxation time ($\sigma/\tau$), and (e) and (f) power factor PF per relaxation time ($S^2\sigma/\tau$) as a function of the chemical potential at temperatures of 300K, and 800K for VTiRhZ (Z=Al, Ga, In)

**Figure. 6:** The lattice thermal conductivity ($\boldsymbol{\kappa_l}$) as a function of the temperature for VTiRhZ (Z=Al, Ga, In) alloys.

**Figure. 7:** (a) and (b) The electronic thermal conductivity ($\boldsymbol{\kappa_e}$) and (c) and (d) figure of merit ($ZT$) as a function of the chemical potential at 300K and 800K for VTiRhZ (Z=Al, Ga, In) alloys.

**List of Tables:**

**Table 1:** The Wyckoff positions 4a (0,0,0), 4c (1/4,1/4,1/4), 4b (1/2,1/2,1/2), 4d (3/4,3/4,3/4) of the atoms in VTiRhZ, (Z= Al, Ga, In) quaternary Heusler alloys for three types of configurations.

**Table 2:** The total energy in eV of VTiRhZ (Z=Al, Ga, In) in the three types of configurations.

**Table 3:** The formation energy $E_{form}$ (eV), optimized lattice constant a (Å), elastic constant $C_{ij}$ (GPa), bulk modulus B (GPa), Young's modulus E (GPa), isotropic shear modulus G (GPa), Poisson's ratios ν, anisotropy factor A, Cauchy pressure $C_p$ (GPa), Pugh's ratio B/G, and melting temperature $T_{melt}$ (K) for the stable type-I structure of VTiRhZ alloys.

**Table 4:** The calculated band gap values $E_g$(eV), spin polarization (P%), total magnetic moment $M_{total}$ (μB) and local magnetic moment per atom (V, Ti, Rh, Z) for VTiRhZ (Z= Al, Ga, In) alloys.

**Table 5:** The Debye temperature $\Theta_D$ (K), average sound velocities $v_m$ (m/s), transverse sound velocities $v_t$ (m/s), longitudinal sound velocities $v_l$ (m/s), density $\rho$ (kg/m3), and Grüneisen parameter $\gamma$

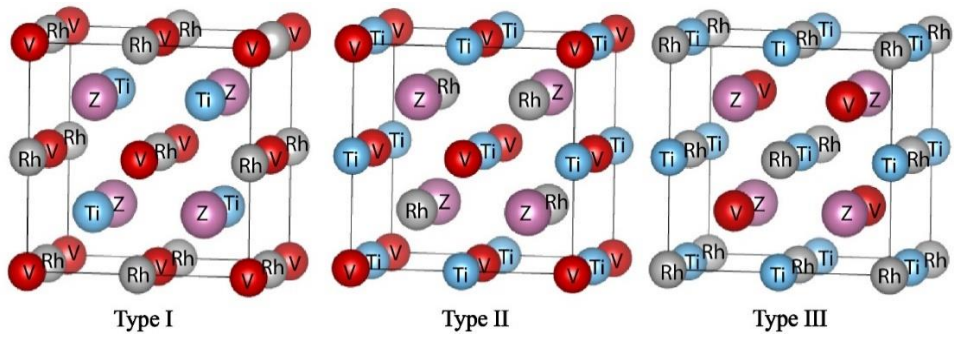

Type I                    Type II                    Type III

**Figure. 1**

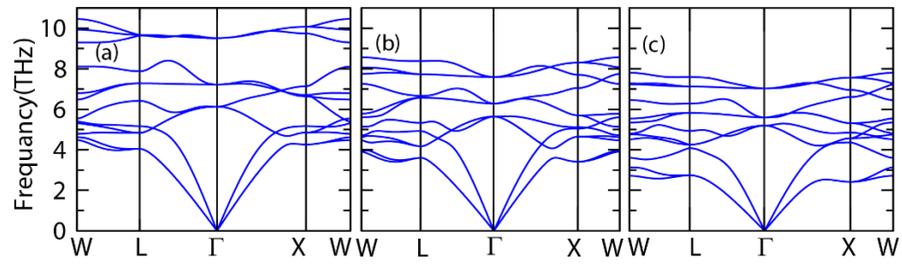

**Figure. 2**

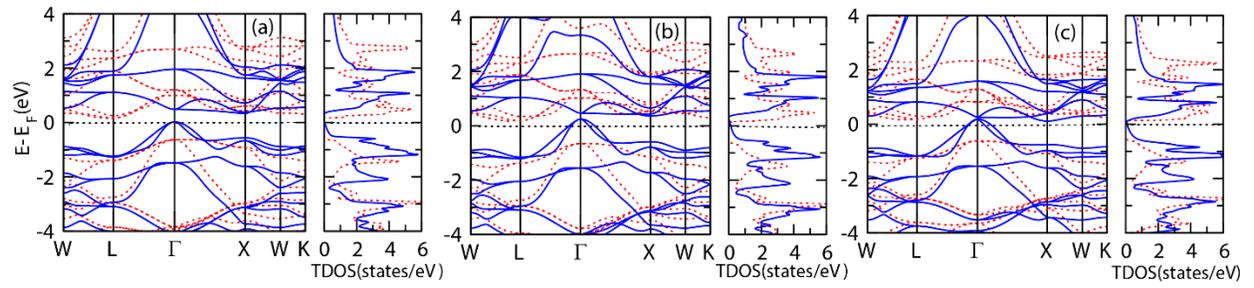

**Figure. 3**

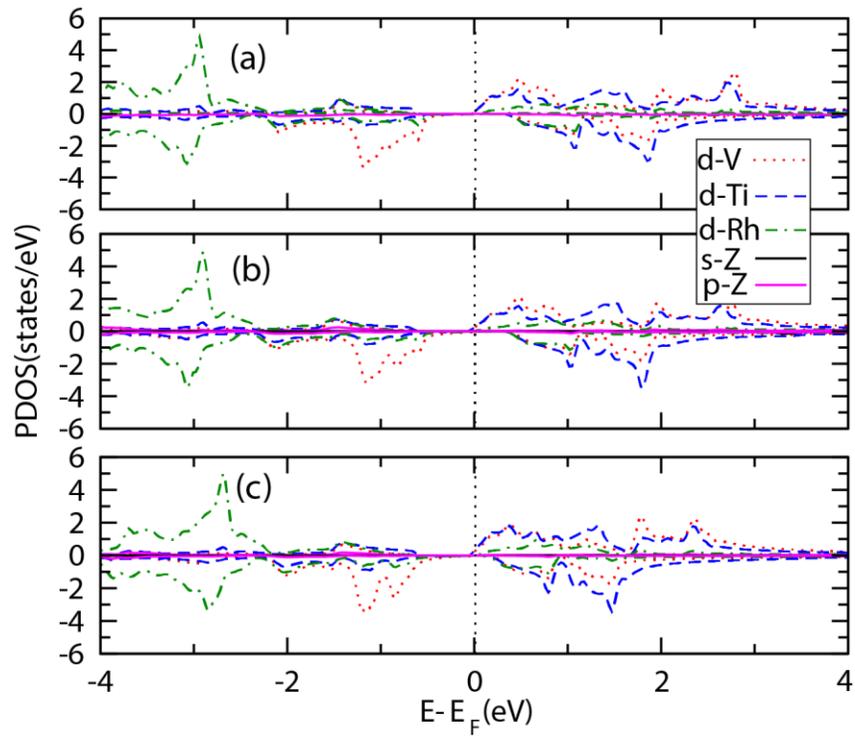

**Figure. 4**

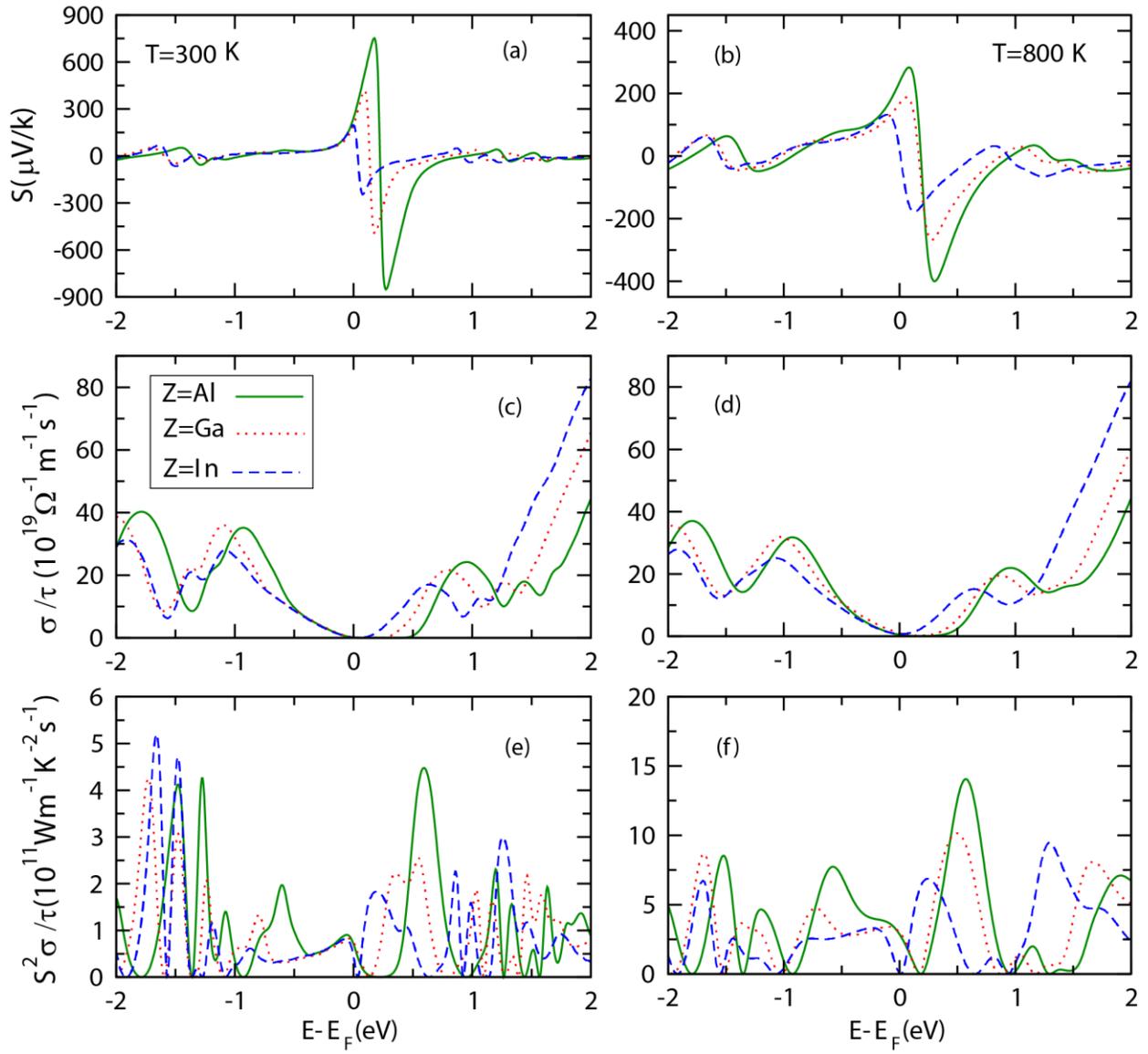

Figure. 5

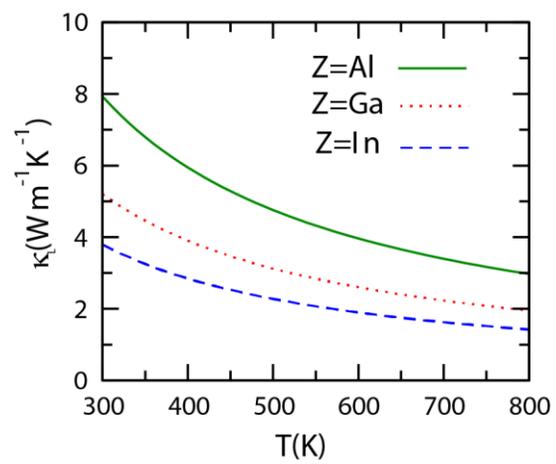

**Figure. 6**

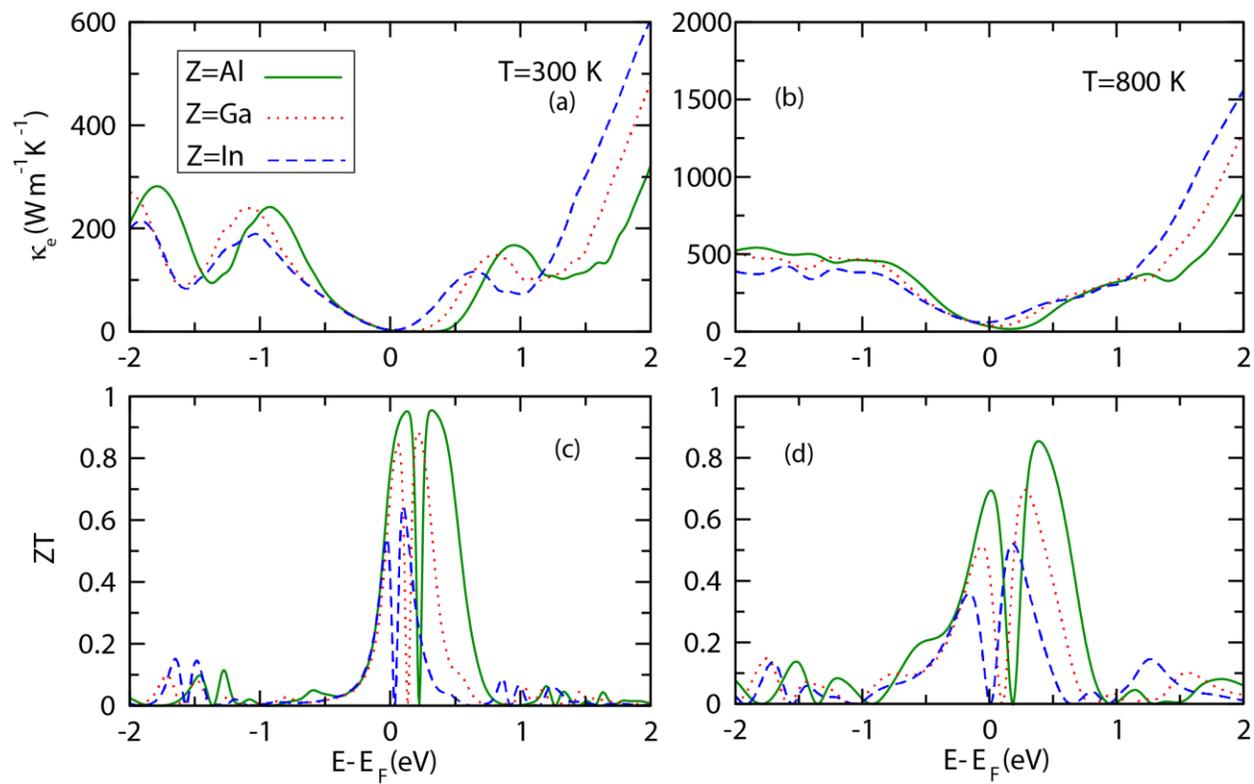

**Figure. 7**

**Table 1**

| Y-type | 4a | 4c | 4b | 4d |
|--------|----|----|----|----|
| **I**   | V  | Ti | Rh | Z  |
| **II**  | V  | Rh | Ti | Z  |
| **III** | Rh | V  | Ti | Z  |

**Table 2**

| Alloys | Type-I | Type-II | Type-III |
|--------|--------|---------|----------|
| **VTiRhAl** | -30.102 | -28.673 | -29.316 |
| **VTiRhGa** | -29.059 | -27.945 | -28.360 |
| **VTiRhIn** | -27.820 | -26.736 | -26.996 |

**Table 3**

| Alloys | $E_{form}$ | a | $C_{11}$ | $C_{12}$ | $C_{44}$ | B | E | G | ν | A | $C_p$ | B/G | $T_{melt}$ |
|--------|-----------|------|--------|--------|--------|-------|-------|------|------|------|-------|------|-------|
| **VTiRhAl** | -2.37 | 6.16 | 297.8 | 103.5 | 86.4 | 168.2 | 236.1 | 93.6 | 0.27 | 0.89 | 211.4 | 1.76 | 2312 |
| **VTiRhGa** | -2.16 | 6.15 | 290.3 | 112.0 | 75.4 | 171.4 | 215.3 | 83.8 | 0.28 | 0.84 | 214.9 | 1.99 | 2268 |
| **VTiRhIn** | -1.28 | 6.38 | 249.7 | 103.4 | 61.9 | 152.1 | 178.7 | 68.7 | 0.29 | 0.84 | 187.8 | 2.15 | 2028 |

**Table 4**

| Alloys | $E_g$ (eV) | | $P$ | $M_{total}$ ($\mu_B$) | $M_V$($\mu_B$) | $M_{Ti}$($\mu_B$) | $M_{Rh}$($\mu_B$) | $M_Z$($\mu_B$) |
|---|---|---|---|---|---|---|---|---|
| | Spin↑ | Spin↓ | | | | | | |
| **VTiRhAl** | 0.04 | 0.62 | 0 | 3.00 | 2.19 | 0.25 | 0.12 | 0.002 |
| **VTiRhGa** | ----- | 0.52 | 100 | 3.00 | 2.21 | 0.28 | 0.12 | -0.01 |
| **VTiRhIn** | ----- | 0.19 | 100 | 3.00 | 2.25 | 0.25 | 0.09 | -0.01 |

**Table 5**

| Alloys | $\Theta_D$ | $v_m$ | $v_t$ | $v_l$ | $\rho$ | $\gamma$ |
|--------|-----------|-------|-------|-------|--------|----------|
| **VTiRhAl** | 514.1 | 4226 | 3802 | 6687 | 6475 | 1.63 |
| **VTiRhGa** | 447.5 | 3674 | 3296 | 6010 | 7714 | 1.77 |
| **VTiRhIn** | 382.8 | 3257 | 2916 | 5449 | 8087 | 1.82 |